\documentclass[12pt,dvips]{article}
\usepackage[dvips]{graphicx}
\usepackage{graphicx}

\graphicspath{{plots/}}
\DeclareGraphicsExtensions{.eps.gz,.eps,.ps,.ps.gz}

\parskip=\itemsep               
\newcommand{\lwig}{\mbox{\,\raisebox{.3ex}
    {$<$}$\!\!\!\!\!$\raisebox{-.9ex}{$\sim$}\,}}
\newcommand{\gwig}{\mbox{\,\raisebox{.3ex}
    {$>$}$\!\!\!\!\!$\raisebox{-.9ex}{$\sim$}}\,}
\newcommand{\lambdabar}{{\hbox{$\lambda_e$\kern-1.9ex\raise+0.45ex\hbox{--}
\kern+0.2ex}}}

\newif\ifhepph

\hepphtrue

\ifhepph\date{\empty}\fi

\title{
\ifhepph{\normalsize\rightline{DESY 02-157}\rightline{hep-ph/0212099}}\fi
\vskip 1cm 
\bf\boldmath
Electroweak instantons/sphalerons at VLHC? 
       \vspace{21mm}} 
\author{
A. Ringwald\\[4mm] 
Deutsches Elektronen-Synchrotron DESY, Hamburg, Germany}

\begin{document}
\begin{titlepage} 
  \maketitle
\begin{abstract}
There is a 
close analogy between electroweak instanton-induced baryon plus lepton number ($B+L$) 
violating processes in Quantum Flavor Dynamics (QFD) and hard QCD instanton-induced chirality 
violating processes in deep-inelastic scattering. 
In view of the recent information about the latter both from lattice simulations and from the 
H1 experiment at HERA, 
it seems worthwhile to reconsider electroweak $B+L$ violation at high energies. 
We present a state of the art evaluation of QFD instanton-induced 
parton-parton cross-sections, as relevant at future high energy colliders in the 
hundreds of TeV regime, such as the projected 
Very Large Hadron Collider (VLHC). 
We find that the cross-sections are unobservably small in a conservative
fiducial kinematical region inferred from the above mentioned QFD--QCD analogy. 
An extrapolation -- still compatible with lattice results and 
HERA -- beyond this conservative limit indicates possible observability 
at VLHC. 
\end{abstract}


\thispagestyle{empty}
\end{titlepage}
\newpage \setcounter{page}{2}

{\em 1.}
The Standard Model of electroweak (Quantum Flavor Dynamics (QFD)) and 
strong (QCD) interactions is remarkably successful and describes quantitatively a wealth
of data accumulated over the last decades. This success is largely based
on the possibility to apply ordinary perturbation theory to the 
calculation of hard, short-distance dominated scattering processes, since
the relevant gauge couplings are small.   

There is, however, a class of processes which 
-- even for small gauge couplings -- can not be described by ordinary
perturbation theory. These processes are associated with axial 
anomalies~\cite{Adler:gk
} and 
manifest themselves as anomalous violation of baryon plus lepton  number ($B+L$) in 
QFD and chirality ($Q_5$) in QCD~\cite{'tHooft:up
}. They are induced by topological fluctuations of the non-Abelian gauge
fields, notably by instantons~\cite{Belavin:fg}.

Such topological fluctuations and the 
anomalous processes induced by them 
are crucial ingredients for an understanding of a number of non-perturbative 
issues in the Standard Model. Indeed, QCD instantons have been argued to play 
an important role in various
long-distance aspects of QCD, such as giving a possible solution to 
the axial $U(1)$ problem~\cite{'tHooft:up
} or being at work in
$SU(n_f)$ chiral symmetry breaking~\cite{Shuryak:1981ff
} (for reviews, see Ref.~\cite{Schafer:1996wv
}). In QFD, on the other hand, 
similar topological fluctuations of the gauge fields and
the associated $B+L$ violating processes are very important at high 
temperatures~\cite{Kuzmin:1985mm
} and have therefore a crucial impact on the evolution of the baryon and lepton
asymmetries of the universe (see Ref.~\cite{Rubakov:1996vz} for a review).

A very interesting, albeit unsolved question is whether manifestations of such topological
fluctuations might be directly observable in high-energy scattering at present or 
future colliders (for 
a short review, see Ref.~\cite{Ringwald:2002iy}). 
This question has been raised originally in the late eighties in 
the context of QFD~\cite{Aoyama:1986ej,Ringwald:1989ee
}. But, despite considerable 
theoretical~\cite{McLerran:1989ab,
Zakharov:1990dj,
Khoze:1991mx,Shuryak:1991pn} and 
phenomenological~\cite{Farrar:1990vb,
Morris:1991bb
} efforts, the actual
size of the cross-sections in the relevant, tens of TeV energy regime was never established (for  
reviews, see Refs.~\cite{Rubakov:1996vz,Mattis:1991bj
}).
Meanwhile, the focus switched to quite similar QCD instanton-induced hard 
scattering processes in deep-inelastic
scattering~\cite{Balitsky:1993jd,Ringwald:1994kr},  
which are calculable from first principles within instanton-perturbation theory~\cite{Moch:1996bs}, 
yield sizeable rates for observable final state signatures  
in the fiducial regime of the 
latter~\cite{Ringwald:1998ek,Ringwald:1999ze,Ringwald:1999jb,Ringwald:2000gt}, 
and are actively searched for at HERA~\cite{Adloff:2002ph}. 
Moreover, larger-size QCD instantons, beyond the semi-classical, instanton-perturbative  
regime, might well be responsible
for the bulk of inelastic hadronic processes and build up soft diffractive 
scattering~\cite{Kharzeev:2000ef
}. 

In view of the close analogy of 
QFD and hard QCD instanton-induced processes in deep-inelastic scattering~\cite{Ringwald:1994kr}, 
emphasized throughout this letter, and of the recent 
information about the latter both from lattice 
simulations~\cite{Smith:1998wt,Ringwald:1999ze,Ringwald:2000gt} and from 
experiment~\cite{Adloff:2002ph}, recalled and elaborated on below,  
it seems worthwhile to reconsider electroweak $B+L$ violation at high energies. 
We therefore present in this letter  
a state of the art evaluation of QFD instanton-induced 
parton-parton cross-sections 
-- quite analogous to the one presented in Ref.~\cite{Ringwald:1998ek} for 
QCD instanton-induced processes --, as relevant at future high energy colliders in the 
hundreds of TeV regime, such as the projected 
Eurasian Long Intersecting
Storage Ring (ELOISATRON)~\cite{eloisatron} or the 
Very Large Hadron Collider (VLHC)~\cite{vlhc}.  
This goes along with a discussion of the implications of the lattice and HERA results -- 
via the above mentioned QFD--QCD analogy -- 
for the fate of electroweak $B+L$ violation in high energy collisions.  
  
\newpage 

{\em 2.}
QCD (QFD) instantons~\cite{Belavin:fg,'tHooft:up
}
are (constrained~\cite{Affleck:1980mp}) minima of the classical Euclidean Yang-Mills action, 
localized in space and Euclidean
time, with unit topological charge (Pontryagin index) $Q=1$. 
In Minkowski space-time, instantons  
describe tunneling transitions between classically degenerate, topologically
inequivalent vacua, differing in their 
winding number (Chern-Simons number) 
by one unit, $\triangle N_{\rm CS}=Q=1$~\cite{Jackiw:1976pf
}. 
The corresponding energy barrier (``sphaleron energy''~\cite{Klinkhamer:1984di}), 
under which the instantons tunnel, is inversely proportional to $\alpha_g\equiv g^2/(4\pi )$, 
the fine-structure constant of the relevant gauge theory,  
and the effective instanton-size $\rho_{\rm eff}$, 
\begin{eqnarray}
\label{barrier}
   M_{\rm sp}\sim \frac{\pi}{\alpha_g\,\rho_{\rm eff}}    \sim
               \cases { \pi \frac{M_W}{\alpha_W}\sim 10\ {\rm TeV} &{\rm in\ QFD~\cite{Klinkhamer:1984di}}\,,\cr
                        {\mathcal Q} & 
{\rm in\ QCD~\cite{Ringwald:1994kr,Moch:1996bs,Ringwald:1998ek}}\,, \cr}
\end{eqnarray}
where ${\mathcal Q}$ is a large 
momentum transfer e.g. in deep-inelastic scattering (DIS), which should be taken $\gwig 10$ GeV in order
to be in the semi-classical, 
instanton-perturbative regime~\cite{Moch:1996bs,Ringwald:1998ek,Ringwald:1999ze,Ringwald:2000gt}.
As mentioned in Sect.~{\em 1}, axial anomalies~\cite{Adler:gk
} force instanton-induced hard scattering processes 
to be always associated with anomalous fermion-number violation~~\cite{'tHooft:up
}, in particular $B+L$ 
violation, $\triangle  B = \triangle L =-n_{\rm gen}\, Q$, 
in the case of QFD  with $n_{\rm gen}=3$ fermion generations, 
and chirality violation, $\triangle Q_5 = 2\,n_f\,Q$, 
in the case of QCD with typically $n_f=3$ light quark flavors.   

\begin{figure}[h]
\begin{center}
 \includegraphics*[width=8cm,bbllx=74pt,bblly=619pt,bburx=314pt,bbury=714,clip=]{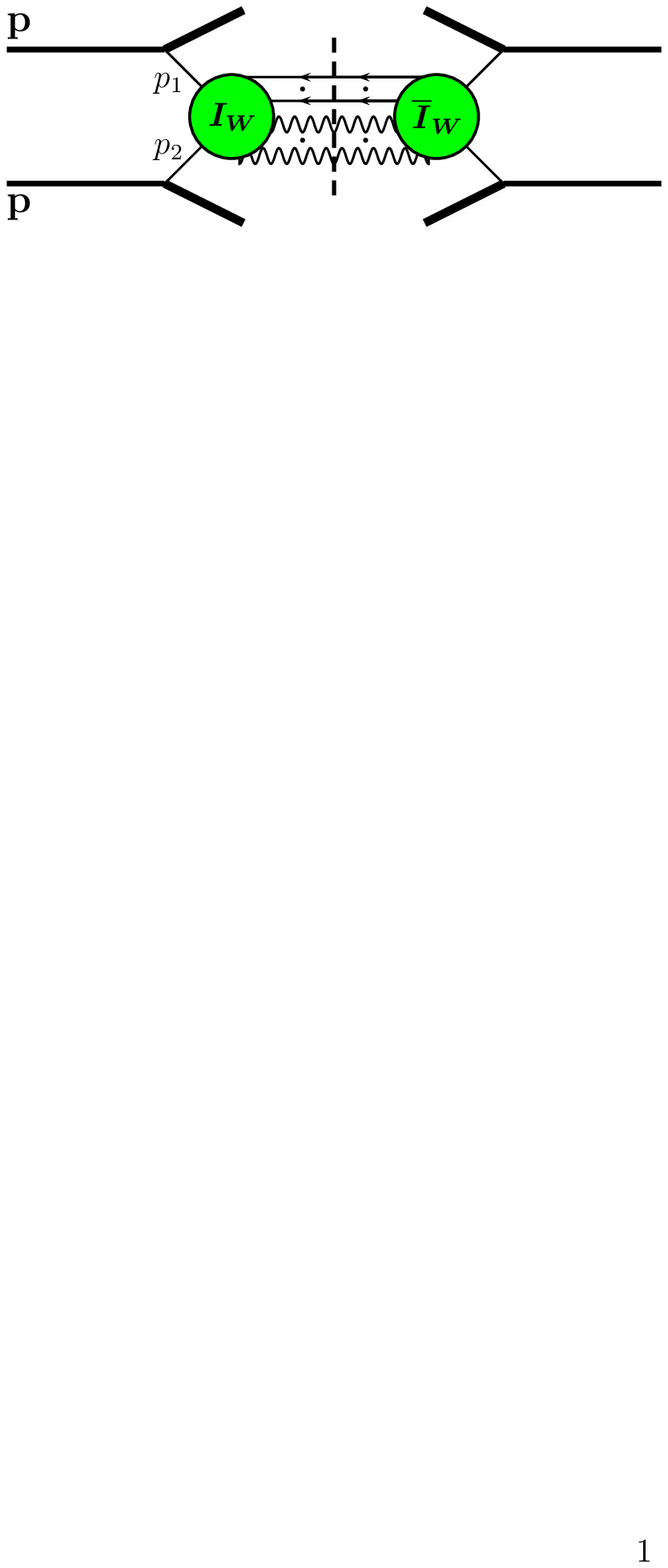}
\hspace{4ex}
 \includegraphics*[width=8cm,bbllx=74pt,bblly=619pt,bburx=314pt,bbury=714,clip=]{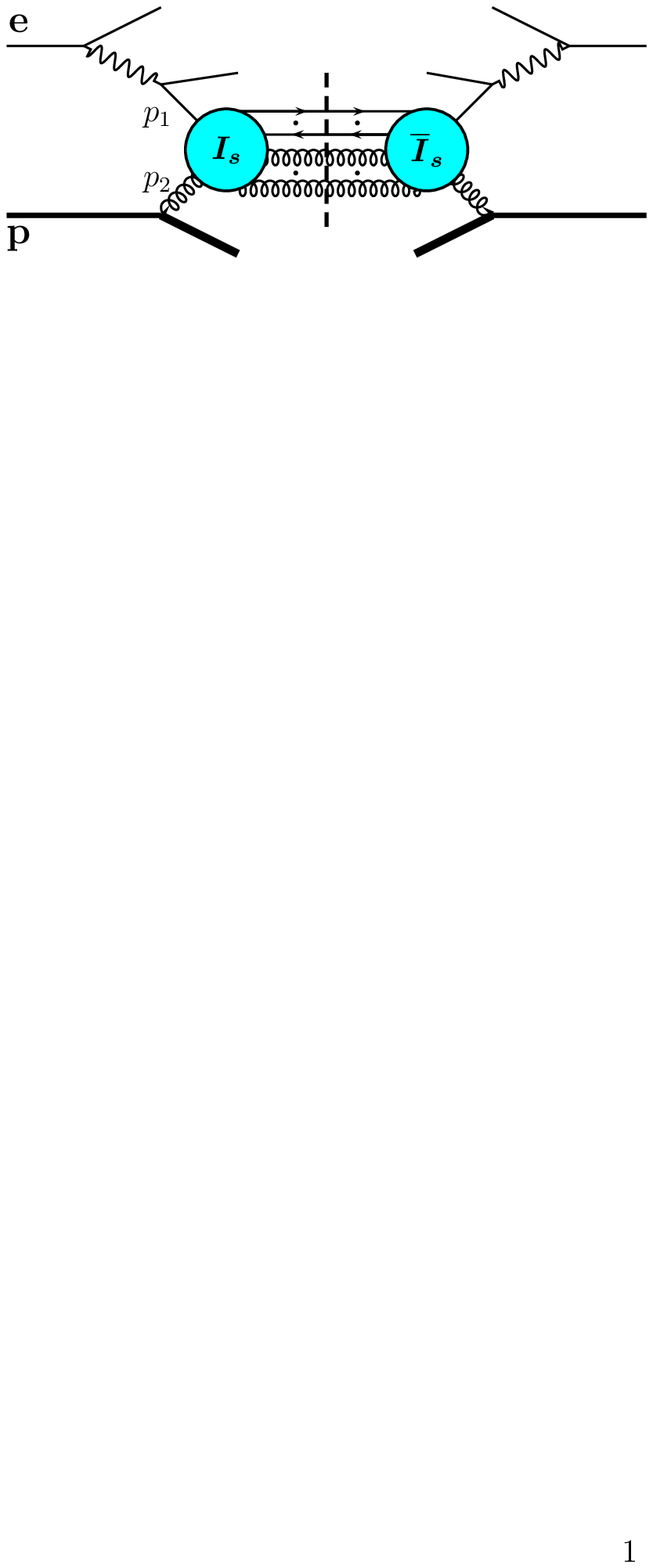}
   \caption[dum]
     {
     Illustration of a QFD instanton-induced process in proton-proton scattering (left) and of a 
     QCD instanton-induced process in deep-inelastic electron-proton scattering (right).
   \label{opt_illu}}
\end{center}
\end{figure}

Instanton-induced total cross-sections for hard parton-parton ($\rm p_1$-$\rm p_2$) 
scattering processes
(cf. Fig.~\ref{opt_illu}) are given in terms of an integral over the 
instanton-anti-instanton\footnote{\label{iai}Both an
instanton and an anti-instanton enter here, 
since we write the cross-section~(\ref{gencross}) as a discontinuity of the $\rm p_1\,p_2$ 
forward elastic scattering amplitude in the
$I\bar I$-background (cf. Fig.~\ref{opt_illu}). 
Alternatively, one may calculate the cross-section by taking the modulus squared of 
amplitudes in the single instanton-background.} ($I\bar I$) collective coordinates 
(sizes $\rho,\bar\rho$, $I\bar I$ distance $R$, relative color 
orientation $U$)~\cite{Ringwald:1998ek} (see also~\cite{Zakharov:1990dj,
Khoze:1991mx,Shuryak:1991pn,Klinkhamer:1991pq,
Balitsky:1992vs})
\begin{eqnarray}
\label{gencross}
      \hat\sigma^{(I)}_{\rm p_1p_2} &\sim &
\frac{1}{2\,p_1\cdot p_2}\,{\rm Im}\,
      \int {\rm d}^4 R\ 
   {\rm e}^{{\rm i}\, ( p_1+p_2 )\cdot R}
\\[1.5ex] \nonumber \mbox{} && \times\ 
      \int\limits_0^\infty {\rm d}\rho 
      \int\limits_0^\infty {\rm d}\overline{\rho}\  
      { D(\rho)\, D(\overline{\rho})}
       \int {\rm d}U\ 
            {\rm e}^{{-\frac{4\pi}{\alpha_g 
                                            }}
      { \Omega\left(U, \frac{R^2}{\rho\bar\rho},\frac{\bar\rho}{\rho}, \ldots \right)}}
\ 
      \left[\omega\left(U, \frac{R^2}{\rho\bar\rho},\frac{\bar\rho}{\rho},\ldots \right)\right]^{n_{\rm fin}}  
\\[1.5ex] \nonumber \mbox{} && \times\  
F\left(\sqrt{-p_1^{2}}\,\rho\right)\,
      F\left(\sqrt{-p_1^{2}}\,\overline{\rho}\right)\,
      F\left(\sqrt{-p_2^2}\,\rho\right)\,F\left(\sqrt{-p_2^2}\,\overline{\rho}\right)
\,{\mathcal P}_{\rm p_1 p_2}^g\left( U, R, \rho, \bar\rho ; p_1\cdot p_2 \right)
\,. 
\end{eqnarray}
Here, the basic blocks arising in instanton-perturbation theory -- the semi-classical expansion
of the corresponding path integral expression about the instanton solution --  
are {\em i)} the instanton-size distribution $D(\rho )$, {\em ii)} the
function $\Omega$, which takes into account the exponentiation of gauge boson 
production~\cite{Ringwald:1989ee,
Zakharov:1990dj
} and can be identified with the 
$I\bar I$-interaction defined via the valley method~\cite{Yung:1987zp,Khoze:1991mx,Verbaarschot:1991sq,%
Arnold:1991dv
}, 
and {\em iii)} the function $\omega$, which summarizes the effects of final-state fermions. Their
number $n_{\rm fin}$ is related to the number $n_{\rm in}$ of initial-state fermions via the anomaly,  
\begin{eqnarray}
\label{nfin}
   n_{\rm fin} + n_{\rm in} \equiv  n_{\rm tot} \equiv  
               \cases { 4\,n_{\rm gen}\   = 12  &{\rm in\ QFD}\,,\cr
                        2\,n_f\       &{\rm in\ QCD}\,. \cr}
\end{eqnarray}
With each initial-state parton $p$, there is an associated ``form
factor''~\cite{Moch:1996bs,Ringwald:1998ek}, 
\begin{equation}
\label{formfac}
F(x)=x\,K_1(x)\ \left\{
\begin{array}{lcllcll}
&\sim& \sqrt{\pi/(2\,x)}\exp(-x) &{\rm \ for\ }&
x& \rightarrow +\infty ,\nonumber\\[2mm]
&=& 1 &{\rm \ for\ }& x&=0\,.
\nonumber\\
\end{array}\right.
\end{equation} 
The function ${\mathcal P}_{\rm p_1 p_2}^g$ in Eq.~(\ref{gencross}) consists of further smooth 
factors~\cite{Ringwald:1998ek}, which will be included in our final result for the special case 
of QFD instantons below. 

{\em Ad i)} The instanton-size distribution $D(\rho )$ is known in 
instanton-perturbation theory, $\alpha_g (\rho^{-1} )\ll 1$,
up to two-loop renormalization group invariance~\cite{'tHooft:up,
Bernard:1979qt,Morris:zi}. In QCD, the loop corrections are sizeable in the phenomenologically interesting
range~\cite{Ringwald:1998ek,Ringwald:1999ze}. However, for our illustrative purposes in this letter 
the one-loop expression for the size distribution,    
\begin{eqnarray}
\label{size-dist}
     { D({\rho})} =
     \frac{d}{\rho^5}
      \left(\frac{2\pi}{\alpha_g (\mu )}\right)^{2\,N_c} 
      (\mu\,\rho )^{\beta_0
      } \ 
     {\rm e}^{-\frac{2\pi}{\alpha_g (\mu )}\,{ S^{(I)}}} \,,    
\end{eqnarray} 
suffices, which, moreover, is numerically adequate for the case of QFD because of its weak coupling, 
$\alpha_W(M_W)\equiv \alpha (M_W)/\sin^2 \hat\theta (M_W)=0.033819(23)$~\cite{Hagiwara:pw}.
In Eq.~(\ref{size-dist}), 
\begin{eqnarray}
\label{beta0}
\beta_0 = 
\frac{11}{3}\,N_c - \frac{1}{6}\,n_s - \frac{1}{3}\,n_{\rm tot} = 
               \cases { 19/6   &{\rm in\ QFD} ($N_c=2, n_s=1, n_{\rm tot}=12$)\,,\cr
                        11 - 2\,n_f/3      &{\rm in\ QCD} ($N_c=3, n_s=0, n_{\rm tot}=2\,n_f$)\,, \cr}
\end{eqnarray} 
denotes the first coefficient in the $\beta$ function, 
\begin{eqnarray}
\label{action}
      { S^{(I)}} =
               \cases {1 + \frac{1}{2}\,M_W^2\,\rho^2 
         +{\mathcal O}\left(M_W^4\,\rho^4 \ln (M_W\,\rho) \right)& 
                     {\rm in\ QFD~\cite{'tHooft:up,
Affleck:1980mp}}\,,\cr
                        1 &{\rm in\ QCD~\cite{Belavin:fg}}\,, \cr}
\end{eqnarray}
the instanton action, $\mu$ the renormalization scale, and $d$ a scheme-dependent constant, which
reads in the $\overline{\rm MS}$ scheme~\cite{Hasenfratz:1981tw
},  
\begin{eqnarray}
\label{dmsbar}
d_{\overline{\rm MS}} = 
\frac{2\,{\rm e}^{5/6}}{\pi^2\,(N_c-1)!(N_c-2)!}\,{\rm
 e}^{-1.511374\,N_c+0.291746\,(n_{\rm tot}+n_s)/2}
\,.
\end{eqnarray}  

\begin{figure}[t]
\vspace{-1.6cm} 
\begin{center}
\includegraphics*[width=8.3cm,clip=]{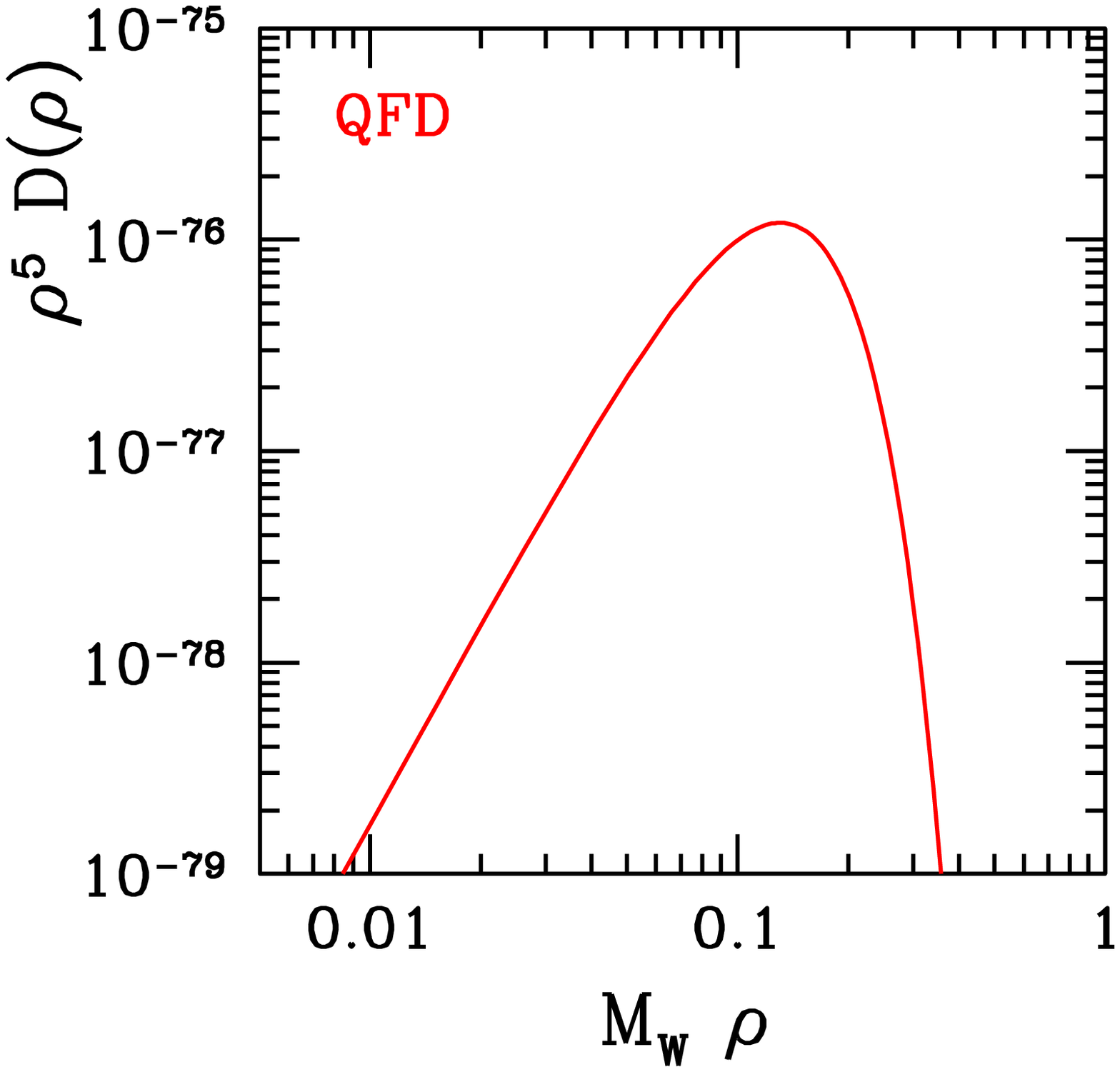}
\hspace{2ex}
\includegraphics*[width=8.3cm,clip=]{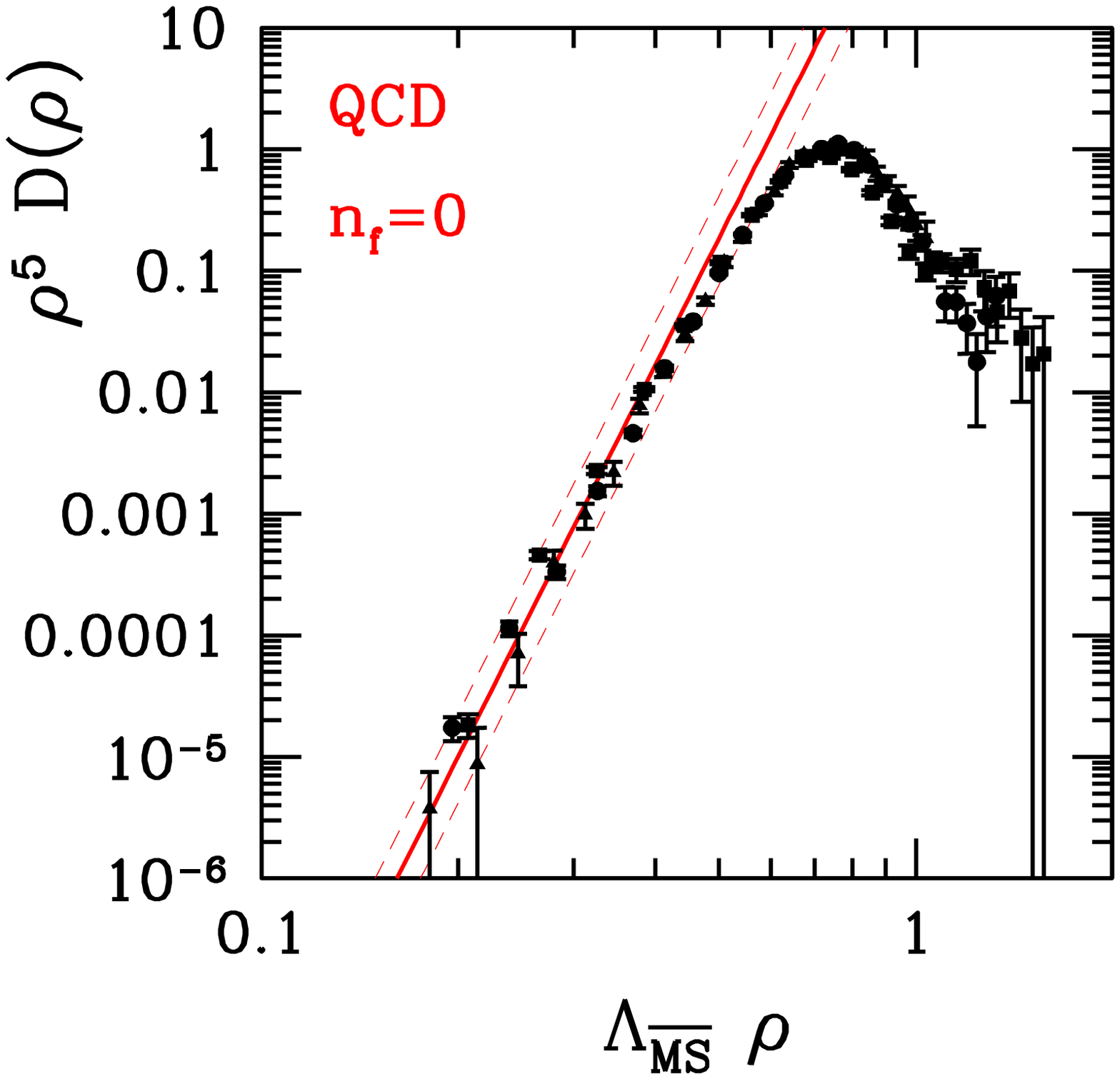}
\vspace{-0.5cm}
\caption[dum]{\label{size-dist-illu}
Instanton-size distributions as predicted in instanton-perturbation theory (solid lines) 
in QFD (left) and quenched ($n_f=0$) QCD (right). 
Both display a powerlike decrease, $\rho^5 D(\rho )\propto \rho^{\beta_0}$, 
towards decreasing sizes due to asymptotic freedom, with $\beta_0=19/6$ for
QFD and $\beta_0=11$ for quenched QCD.
For QCD (right), the two-loop renormalization group invariant prediction for the size distribution from 
Ref.~\cite{Morris:zi} together with the 3-loop form of $\alpha_{\overline{\rm MS}}$, with
$\Lambda^{(0)}_{\overline{\rm MS}}=238\pm 19$~MeV from the ALPHA collaboration~\cite{Capitani:1998mq}, was used.
The error band (dashed lines) results from the errors in $\Lambda_{\overline{\rm MS}}$ and a variation
of $\mu = 1\div 10$~GeV.  
{\em Left:} Towards large sizes, 
$\rho > \rho_{\rm max}=0.13/M_W$, 
the QFD instanton size distribution decreases
exponentially due to the Higgs mechanism.
{\em Right:} 
For large sizes, $\Lambda_{\overline{\rm MS}}\,\rho \gwig 0.75$, the QCD instanton size 
distribution, as determined from recent high-quality lattice data from UKQCD~\cite{Smith:1998wt}$^{{\ref{lattice-ref}}}$, 
appears to decrease exponentially, $\propto\exp (-c\rho^2)$~\cite{Ringwald:1999ze,Shuryak:1999fe}, 
similar to the QFD size distribution (left), but unlike the instanton-perturbative prediction (solid).  
For $\Lambda_{\overline{\rm MS}}\, \rho\lwig 0.42$, on the other hand, one observes
a remarkable agreement with the predictions from  
instanton-perturbation theory (solid)~\cite{Ringwald:1999ze,Ringwald:2000gt}. 
}
\end{center}
\end{figure}

The validity of instanton-perturbation theory, on which the prediction of the instanton-induced
subprocess cross-section~(\ref{gencross}) is based, requires instantons of small enough size,
$\alpha_g (\rho^{-1})\ll 1$. 
In QFD, this is guaranteed by the exponential decrease 
$\propto\exp (-\pi\,M_W^2\,\rho^2/\alpha_W )$ (cf.~(\ref{action})) 
of the size distribution~(\ref{size-dist}) for 
$\rho > \rho_{\rm max}\equiv \sqrt{\beta_0\,\alpha_W/(2\pi)}/M_W=0.13/M_W$ (cf. Fig.~\ref{size-dist-illu} (left)). 
Therefore, the relevant contributions to the
size integrals in~(\ref{gencross}) arise consistently from the perturbative region 
($\alpha_W(\rho^{-1})\ll 1$) even if both initial partons are on-shell, $p_1^2\approx p_2^2\approx 0$, as 
relevant for electroweak instanton-induced processes in proton-proton scattering at VLHC (cf. Fig.~\ref{opt_illu} (left)).

In QCD, on the other hand, the perturbative expression~(\ref{size-dist}) for the size distribution
has a power-law behavior, $\propto \rho^{\beta_0 -5}$ (cf. Fig.~\ref{size-dist-illu} (right)). The latter 
generically causes the dominant contributions to observables like the cross-section~(\ref{gencross}) 
to originate from large $\rho\sim \Lambda^{-1}\Rightarrow \alpha_s(\rho^{-1})\sim 1$ and thus
often spoils the applicability of instanton-perturbation theory.
Deep-inelastic scattering, however, offers a unique possibility to probe 
the predictions of instanton-perturbation 
theory~\cite{Moch:1996bs,Ringwald:1998ek,Ringwald:1999ze,Ringwald:1999jb,Ringwald:2000gt}. This can be 
understood as follows.   
In deep-inelastic electron-proton scattering, the virtual photon splits into a quark and an 
antiquark, one of which, $p_1$ say, enters the instanton subprocess (cf. Fig.~\ref{opt_illu} (right)).
This parton carries a space-like virtuality ${\hat Q}^{2}\equiv -p_1^2\geq 0$, which can be
made very large by kinematical cuts on the final state. In this case the contribution of large
instantons to the integrals is suppressed by the exponential form 
factors~(\ref{formfac}) in expression~(\ref{gencross}), $\propto {\rm e}^{-\hat Q (\rho+\bar\rho )}$,    
and instanton-perturbation theory becomes exploitable, i.e. predictive~\cite{Moch:1996bs,Ringwald:1998ek}. 
In this connection it is quite welcome that 
lattice data on the instanton content of the quenched ($n_f=0$) 
QCD vacuum~\cite{Smith:1998wt}\footnote{\label{lattice-ref}For further, qualitative similar lattice data, 
see Refs.~\cite{Hasenfratz:1998qk,GarciaPerez:1998ru} and the 
reviews~\cite{Negele:1998ev,
Stamatescu:2000ch}.}
can be used to infer the region of validity of instanton-perturbation
theory for $D(\rho )$~\cite{Ringwald:1999ze,Ringwald:2000gt}: 
As illustrated in Fig.~\ref{size-dist-illu} (right), there is very good agreement 
for $\Lambda_{\overline{\rm MS}}\,\rho\lwig 0.42$. 

{\em Ad ii)} A second important building block of the cross-section~(\ref{gencross}) is the function
$\Omega(U,R^2/(\rho\overline{\rho}),\overline{\rho}/\rho)$, appearing
in the exponent with a large numerical 
coefficient $4\pi/\alpha_g$. It incorporates the effects of final-state
(gauge) bosons, mainly $W$'s and $Z$'s in the case of QFD and gluons in the case of 
QCD. Within strict instanton-perturbation theory, it is given in form
of a perturbative expansion~\cite{Zakharov:1990dj,
Arnold:1991dv,
Ringwald:1998ek} for large $I\bar I$-distance $R^2$. 
Beyond this expansion, one may identify $\Omega$ with the interaction between an instanton and 
an anti-instanton,  
which may be systematically evaluated by means of  
the so-called $I\bar I$-valley method~\cite{Yung:1987zp}. 
The corresponding interaction has been found analytically for the case of 
pure $SU(2)$ gauge theory\footnote{\label{emb}For the embedding 
of the $SU(2)$ $I\bar I$-valley into $SU(3)$, 
see e.g. Ref.~\cite{Ringwald:1999ze}.}~\cite{Khoze:1991mx,Verbaarschot:1991sq},   
\begin{equation}
\Omega_g
=
\Omega_0 + \Omega_1\,u_0^2 +\Omega_2\,u_0^4\,,
\label{valley}
\end{equation}
with
\begin{eqnarray}
\nonumber
\Omega_0 &=& 2\,\frac{z^4-2z^2+1+2\, (1-z^2)\ln z}{(z^2-1)^3}\,,
\\[1.5ex] \label{valley-int}
\Omega_1 &=& -8\,\frac{z^4-z^2+(1-3\,z^2)\ln z}{(z^2-1)^3}\,,
\\[1.5ex] \nonumber
\Omega_2 &=& -16\,\frac{z^2-1-(1+z^2)\ln z}{(z^2-1)^3}\,.
\end{eqnarray}
Due to conformal invariance of classical pure Yang-Mills theory, it depends on the 
sizes $\rho$, $\bar\rho$, and the $I\bar I$-distance $R$ only through the 
``conformal separation'', 
\begin{equation}
\label{conf-sep}
z = \frac{1}{2} \left( \xi+\sqrt {{\xi}^{2}-4}\right)\,, \hspace{9ex}
\xi = \frac{R^2}{\rho\overline{\rho}}+\frac{\overline{\rho}}{\rho}+
\frac{\rho}{\overline{\rho}}\,\geq\, 2\,, 
\end{equation} 
and on the relative color orientation$^{\ref{emb}}$ $U=u_0+{\rm i}\,\sigma^k u_k$, with $u_0^2+u^k u_k=1$,  
only through $u_0$. 

\begin{figure}[t]
\vspace{-1.5cm}  
\begin{center}
\includegraphics*[width=7.6cm,clip=]{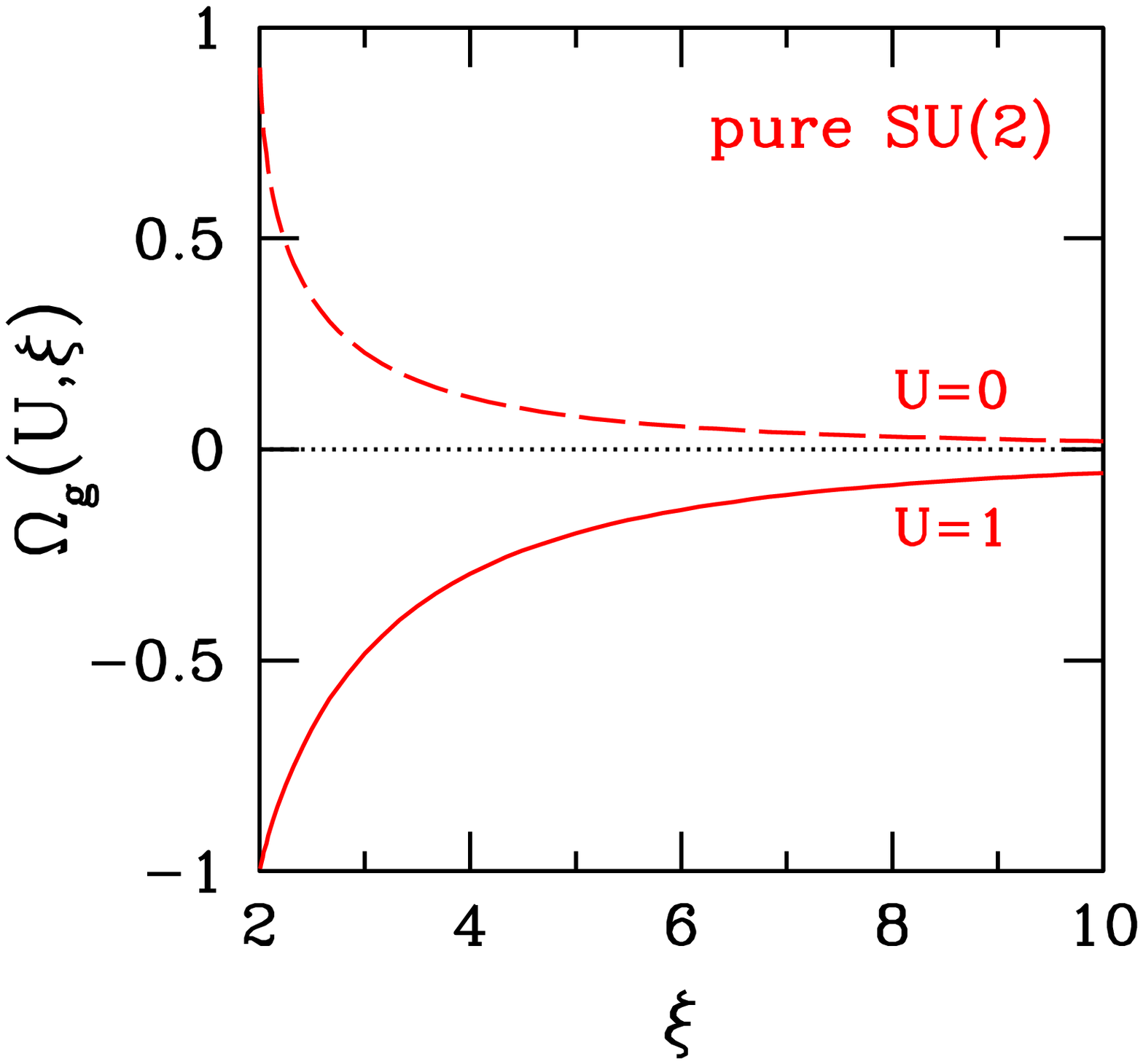}
\hspace{2ex}
\includegraphics*[width=6.4cm]{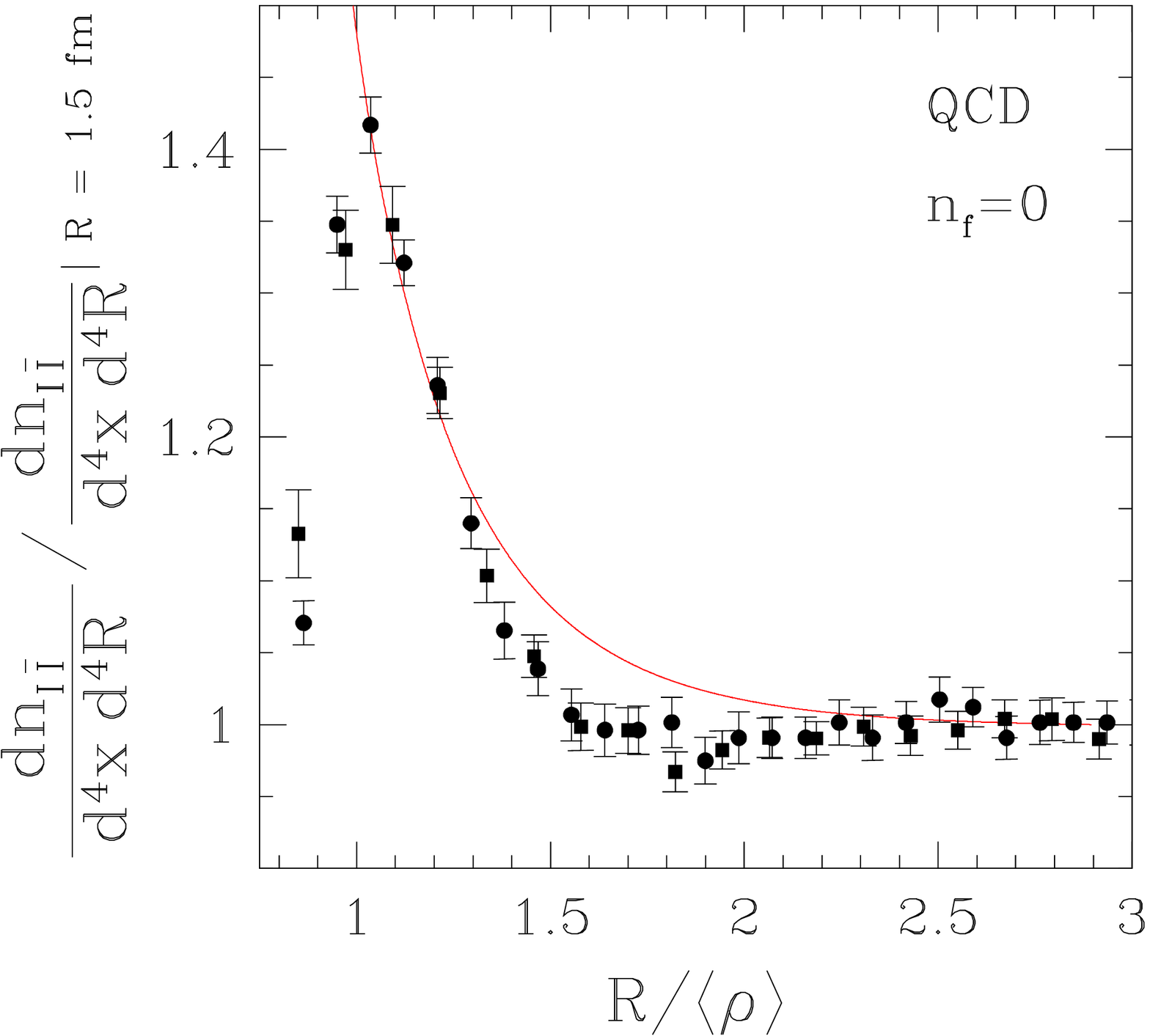}
\vspace{-2ex}
\caption[dum]{\label{qcd-lattice-dist}
{\em Left:} $I\bar I$-valley interaction~(\ref{valley}) as function of 
conformal separation $\xi$~(\ref{conf-sep}) for the most attractive 
relative orientation ($U=1$, solid) and the most repulsive relative orientation 
($U=0$, dashed).
{\em Right:}  
Illustration of the agreement
of recent high-quality lattice data~\cite{Smith:1998wt} 
for the $I\overline{I}$-distance
distribution with the predictions from  
instanton-perturbation theory (solid) 
for $R/\rho\gwig 1.05$~\cite{Ringwald:1999ze,Ringwald:2000gt}. 
}
\end{center}
\end{figure}

Note that $I\bar I$-pairs with the most attractive relative orientation,  $U=1$,  
give the dominant contribution to the cross-section~(\ref{gencross}) in the weak coupling
regime, $\alpha_g\ll 1$. For this relative orientation, the $I\bar I$-valley represents 
a gauge field configuration of steepest descent interpolating between an 
infinitely separated $I\bar I$-pair, corresponding to twice the instanton action, 
$S^{(I\bar I)}=2\,[1+\Omega_g(U=1,\xi=\infty)]=2$, and a strongly overlapping one, 
annihilating to the perturbative vacuum at $\xi = 2$ ($R=0,\rho=\bar\rho$), 
corresponding to vanishing action $S^{(I\bar I)}=2\,[1+\Omega_g(U=1,\xi=2)]=0$ 
(cf. Fig.~\ref{qcd-lattice-dist} (left)).
It is thus clear that near $\xi\approx 2$ the semi-classical approximation 
based on the $I\bar I$-valley breaks down and no reliable non-perturbative information can be extracted 
from it. 

Here again high-quality lattice data~\cite{Smith:1998wt} on the $I\bar I$-distance distribution 
in quenched QCD allow to estimate the fiducial region in  $\xi$   
or more specifically in $R/\langle\rho\rangle$, where $\langle \rho\rangle\approx 0.5$~fm 
is the average instanton/anti-instanton size measured on the lattice (cf. Fig.~\ref{size-dist-illu} (right)).    
One finds good agreement with the predictions from instanton-perturbation theory for 
$R/\langle\rho\rangle\gwig 1.0\div 1.05$~\cite{Ringwald:1999ze,Ringwald:2000gt} 
(cf. Fig.~\ref{qcd-lattice-dist} (right)). 
In this case, however, there are remaining ambiguities. 
{\em a)} The integrations over $\rho$, $\bar\rho$ 
in the $I\bar I$-distance distribution ${\rm d}n_{I\bar I}/({\rm d}^4x\, {\rm d}^4 R  )$ imply
significant contributions also from larger instantons with 
$0.42\,\lwig\, \Lambda_{\overline{\rm MS}}\,\rho, \Lambda_{\overline{\rm MS}}\,\overline{\rho}\,\lwig\, 1$, 
outside the region of instanton-perturbation theory. 
A more differential lattice measurement of the distance distribution, 
${\rm d}n_{I\bar I}/({\rm d}^4x\, {\rm d}^4 R\, {\rm d}\rho\, {\rm d}\bar\rho )$,
which includes also differentials with respect to the sizes $\rho$ and $\bar\rho$, 
and eventually a test of its conformal properties would resolve these theoretical ambiguities.   
{\em b)} 
Furthermore, at small $I\bar I$-separation $R<(\rho +\bar\rho)/2$, the extraction of the 
$I\bar I$-distance distribution from the quenched QCD lattice data is quite ambiguous since there is no 
principal distinction between a trivial gauge field fluctuation and an $I\bar I$-pair at small 
separation. 
This is reflected in a considerable dependence on the cooling method/amount used to 
infer properties of the $I\bar I$-distance distribution~\cite{GarciaPerez:1998ru,Stamatescu:2000ch}. 
A simple extrapolation of lattice results on the topological structure of 
quenched $SU(2)$ gauge theory~\cite{GarciaPerez:1998ru} to zero ``cooling radius'' indicates  
$\langle R/(\rho+\bar\rho)/2\rangle\approx 0.5$, i.e. strongly overlapping $I\bar I$-pairs in the vacuum, 
unlike Fig.~\ref{qcd-lattice-dist} (right).   
Therefore, the fiducial region $R^2/(\rho\bar\rho)\geq 1$ for the reliability of instanton-perturbation
theory inferred from lattice data should be considered as quite conservative.

{\em Ad iii)} Finally, as the last important building block of~(\ref{gencross}), let us just quote
the result from Ref.~\cite{Ringwald:1998ek} for the fermionic overlap 
integral $\omega$~\cite{Shuryak:1991pn}  
in the most attractive relative orientation, $U=1$, 
\begin{equation}
      \label{exactom}
      \omega =
      \frac{3\pi}{8}\,\frac{1}{z^{3/2}}\ 
      \phantom{}_2F_1\left(\frac{3}{2},\frac{3}{2};4;1-\frac{1}{z^2}\right)\,.
\end{equation}

{\em 3.}
In the weak-coupling regime, $\alpha_g\ll 1$, the collective coordinate integrals in 
the cross-section~(\ref{gencross}) can be performed in the saddle-point approximation, 
where the relevant effective exponent reads\footnote{\label{qcd-mod}In the case of QCD, 
some additional terms, which 
arise from the running of $\alpha_s$ and are formally of pre-exponential nature, have to be 
included in 
Eqs.~(\ref{holy-exp}) and (\ref{saddle-coll}) for 
numerical accuracy~\cite{Ringwald:1998ek}. In this letter, we adopt the simplified 
expressions~(\ref{holy-exp}),  
(\ref{saddle-coll})
which suffice for illustrative purposes and are numerically adequate for QFD.}
\begin{eqnarray}
\label{holy-exp}
-\Gamma &\equiv & {\rm i}\, ( p_1+p_2 )\cdot R
\\[1.5ex] \nonumber  && -   
               \cases {  
   \frac{4\pi}{\alpha_W(\mu )}
       \left[ 1 + \frac{1}{4}\, M_W^2 (\rho^2+\bar\rho^2 )
       + \Omega_g \left(U, \frac{R^2}{\rho\bar\rho}, \frac{\bar\rho}{\rho}  \right) \right] 
     &{\rm in\ QFD} ($p_1^2=p_2^2=0$)\,,\cr
  \hat Q\,(\rho +\bar\rho ) + \frac{4\pi}{\alpha_s(\mu )}
       \left[ 
       1
       + \Omega_g \left(U, \frac{R^2}{\rho\bar\rho}, \frac{\bar\rho}{\rho}  \right) \right]
      &{\rm in\ QCD} (DIS: $-p_1^2=\hat Q^{2}>0, p_2^2=0$)\,. \cr}
\end{eqnarray}       
For the case of QFD, we have neglected in~(\ref{holy-exp}) the Higgs part $\Omega_h$ of the 
$I\bar I$-interaction and took for the gauge part the one from the pure gauge theory, $\Omega_g$.    
This should be reliable as long as the dominant contribution to the QFD instanton-induced cross-section
is due to the multiple production of transverse $W$'s and $Z$'s -- as is the case at energies below 
the sphaleron~(\ref{barrier}) -- rather than of longitudinal ones and of Higgs bosons~\cite{Khoze:1991mx}.  
%
\begin{figure}
\vspace{-2cm}  
\begin{center}
\includegraphics*[width=10.2cm]{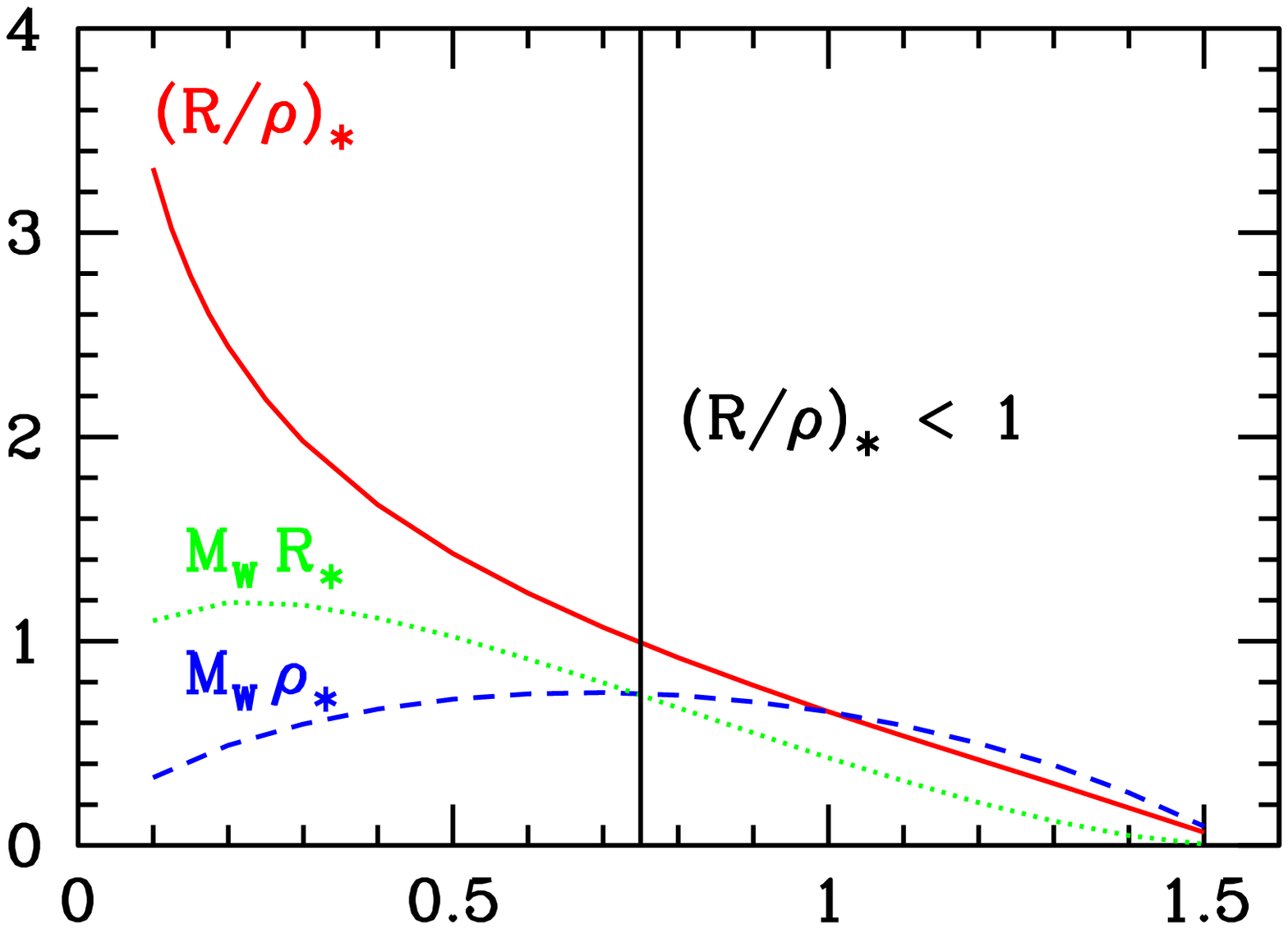}
\mbox{}\vspace{-4cm}
\includegraphics*[width=10.2cm]{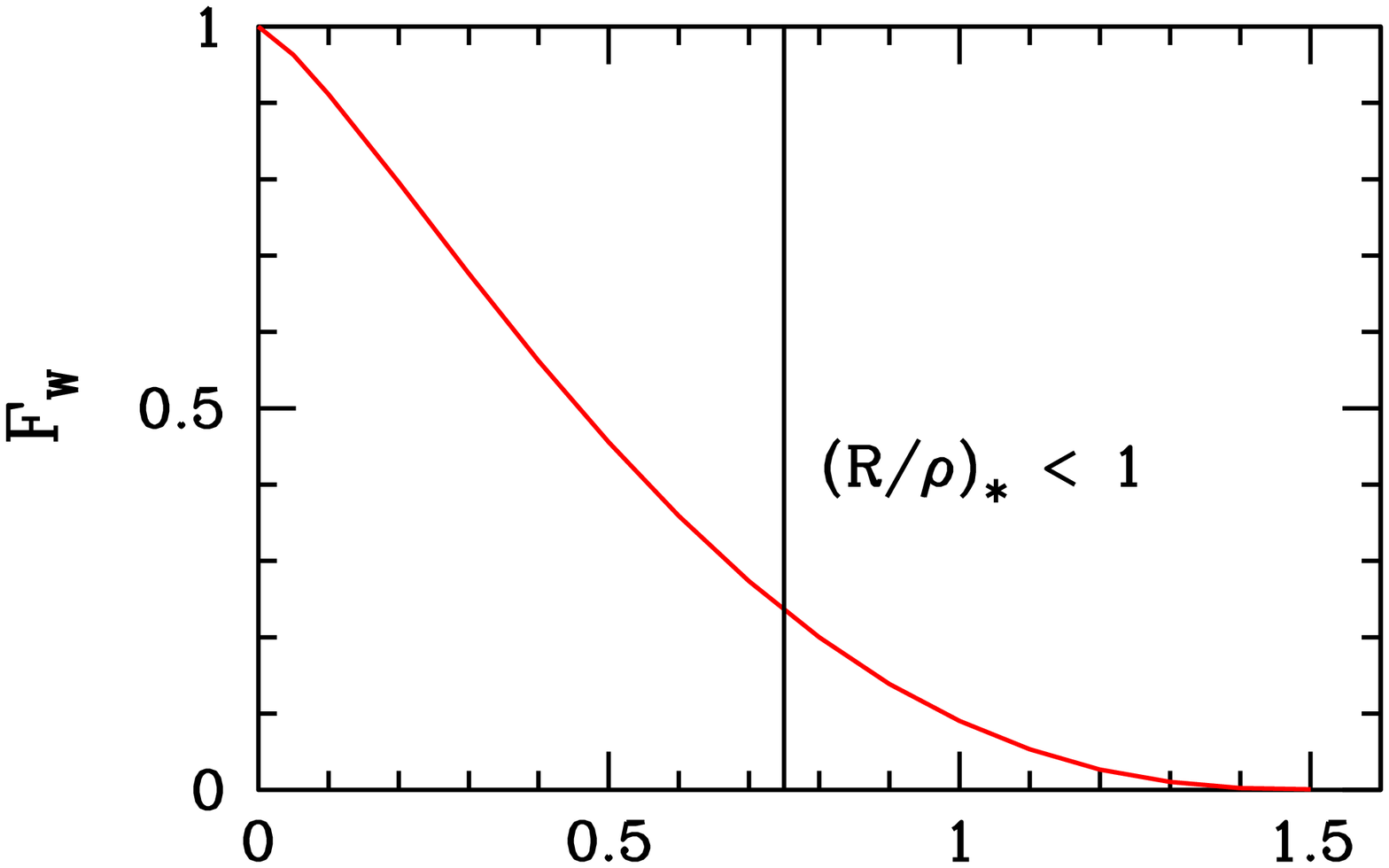}
\mbox{}\vspace{-4cm}
\includegraphics*[width=10.2cm]{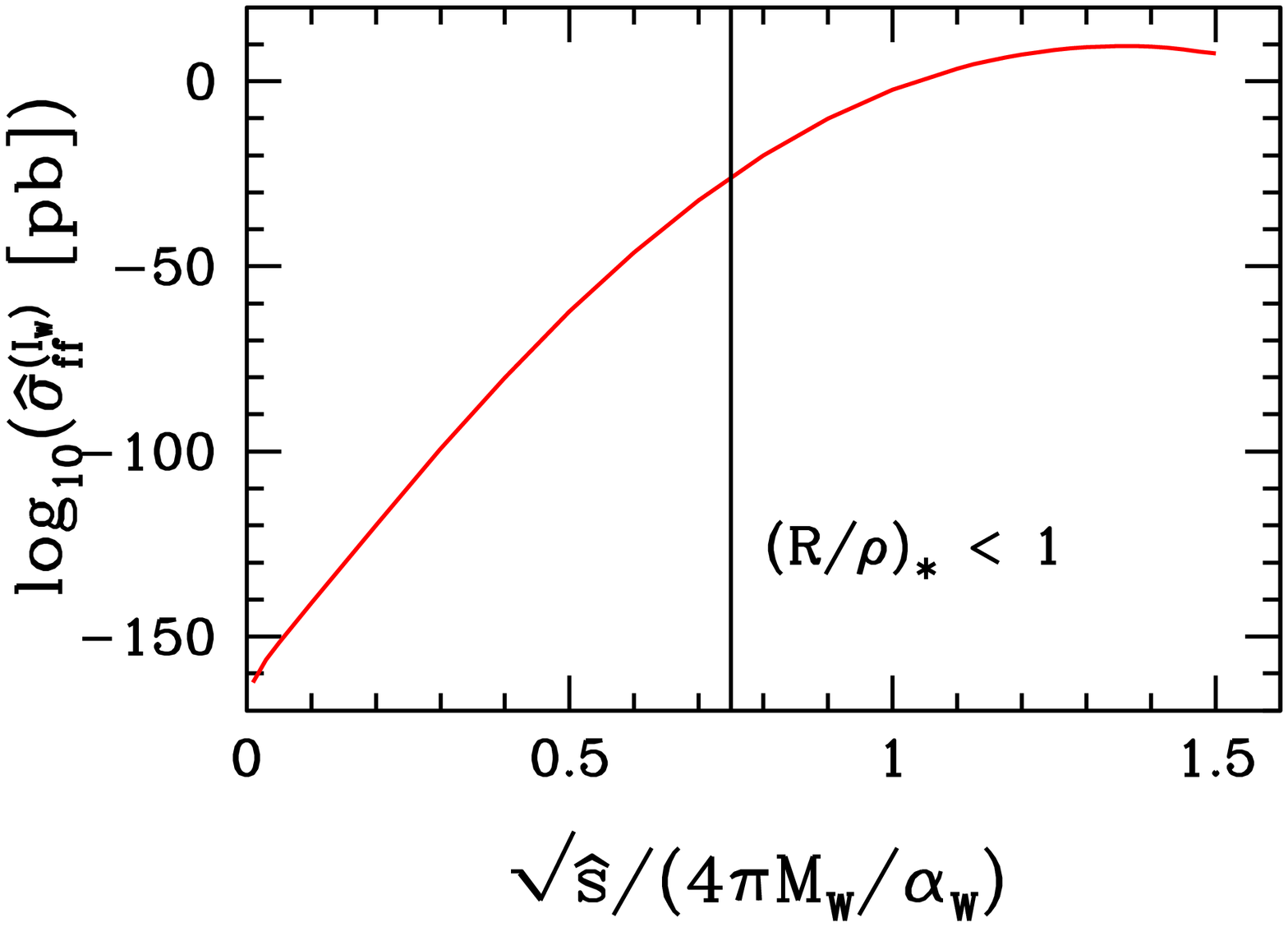}
\vspace{-1cm}
\caption[dum]{\label{saddle-holy}
QFD instanton subprocess cross-section related quantities, 
as function of scaled parton-parton center-of-mass energy $\sqrt{\hat s}/(4\pi M_W/\alpha_W )$.
{\em Top:} Saddle point values for collective coordinates~\cite{Khoze:1991mx}. 
{\em Middle:} Holy-grail function, $\hat\sigma^{(I_W)}\propto \exp [-(4\pi/\alpha_W)\,F_W ]$~\cite{Khoze:1991mx}.
{\em Bottom:} Total cross-section $\hat\sigma^{(I_W)}_{\rm ff}$ 
for QFD instanton-induced fermion-fermion scattering, ${\rm f+f}\stackrel{I_W}{\to}{\rm all}$.   
}
\end{center}
\end{figure}
%
The saddle-point equations, $\partial\Gamma/\partial\chi_{\mid \chi_\ast}=0$, 
with $\chi = \{ U,R,\rho,\bar\rho\}$, 
following from~(\ref{holy-exp}) imply $U_\ast =1$, $\rho_\ast = \bar\rho_\ast$, and
can be summarized as$^{{\ref{qcd-mod}}}$ 
\begin{eqnarray}
\label{saddle-coll}
               \cases {  
\left(\frac{R}{\rho}\right)_\ast = M_W\rho_\ast\,\left( \frac{4\pi M_W/\alpha_W}{\sqrt{\hat s}}\right) 
\,,\hspace{1ex} 
\frac{1}{2}\, \left( M_W\rho_\ast\right)^2 =
\left[(\xi_\ast -2 )\,
\frac{\partial}{\partial\xi_\ast} \Omega_g (1,\xi_\ast )\right]_{\mid \xi_\ast =2+\left(\frac{R}{\rho}\right)_\ast^2}
     &{\rm in\ QFD} \,,\cr
\left(\frac{R}{\rho}\right)_\ast = 2\,\frac{\hat Q}{\sqrt{\hat s}}\,,\hspace{1ex}  
\hat Q\,\rho_\ast = \frac{4\pi}{\alpha_s}\,\left[(\xi_\ast -2 )\,
\frac{\partial}{\partial\xi_\ast} \Omega_g (1,\xi_\ast )\right]_{\mid \xi_\ast =2+\left(\frac{R}{\rho}\right)_\ast^2}
      &{\rm in\ QCD} \,, \cr}
\end{eqnarray}   
where $(p_1+p_2)^2=\hat s$ denotes the parton-parton center-of-mass (cm) energy. 
To exponential accuracy, the cross-section~(\ref{gencross}) is then given by
\begin{eqnarray}
\label{gen-cross-holy}
\hat\sigma^{(I)} \propto 
{\rm e}^{-\Gamma_\ast} \equiv
{\rm e}^{-\frac{4\pi}{\alpha_g}\,F_g(\epsilon )}
\,,
\end{eqnarray}
where
\begin{eqnarray} 
\epsilon &=& 
               \cases {  
\frac{\sqrt{\hat s}}{4\pi M_W/\alpha_W}
     &{\rm in\ QFD} \,,\cr
\frac{\sqrt{\hat s}}{\hat Q}
      &{\rm in\ QCD} \,, \cr}
\\[1.5ex]  
\label{holy-grail}
F_g  &=& 
               \cases {  
\left[1+\Omega_g (1,\xi_\ast )
-(\xi_\ast -2 )\,
\frac{\partial}{\partial\xi_\ast} \Omega_g (1,\xi_\ast )
\right]_{\mid \xi_\ast =2+\left(\frac{R}{\rho}\right)_\ast^2}
     &{\rm in\ QFD} \,,\cr
\left[1+\Omega_g (1,\xi_\ast )\right]_{\mid \xi_\ast =2+\left(\frac{R}{\rho}\right)_\ast^2}
      &{\rm in\ QCD} \,. \cr}
\end{eqnarray}
Both in QFD as well as in QCD, the prediction~(\ref{holy-grail}) for the 
``holy-grail function~\cite{Mattis:1991bj
}''  $F_g(\epsilon )$ decreases monotonically for increasing scaled energy $\epsilon$ from $F_g(0)=1$. It
approaches zero, $F_g\to 0$, at asymptotic energies, $\epsilon\to\infty$ (cf. Fig.~\ref{saddle-holy} (middle)). 
Thus, at all finite energies, the cross-section~(\ref{gen-cross-holy}) is formally 
exponentially suppressed and there is  no apparent problem with unitarity~\cite{Rubakov:1996vz}.  
On the other hand, it is seen that at high cm energies the $I\bar I$-interaction is probed 
at small distances, $(R/\rho)_\ast\sim 1$ (cf. Fig.~\ref{saddle-holy} (top)), making
the semi-classical and saddle-point evaluation unreliable.    
In this connection, the information on the fiducial region in 
$R/\langle\rho\rangle$ of the instanton-perturbative description from QCD lattice simulations 
(cf. Fig.~\ref{qcd-lattice-dist} (right)) can be most appreciated. 
Note furthermore that, in the case of QFD, $M_W R_\ast\,\lwig\, 1$ in the whole
energy range considered in Fig.~\ref{saddle-holy} (top), justifying a posteriori the 
approximation of the full valley interaction in QFD by the one from the pure gauge theory, $\Omega_g$.  

Further information on the fiducial region in $(R/\rho)_\ast$ may be obtained from
DIS experiments at HERA. Meanwhile, the results of a first dedicated search for QCD instanton-induced processes 
in DIS  have been published by the H1 collaboration~\cite{Adloff:2002ph}. In this study,   
the theory and phenomenology of hard 
QCD instanton-induced processes in DIS developed by Fridger Schrempp and 
myself~\cite{Ringwald:1994kr,Moch:1996bs,Ringwald:1998ek,Ringwald:1999ze,Ringwald:1999jb,Ringwald:2000gt} 
has been used heavily. 
Several observables characterising the hadronic final state of QCD instanton-induced
events were exploited to identify a potentially instanton-enriched domain. The results obtained
are intriguing but non-conclusive. While an excess of events with instanton-like topology
over the expectation of the standard DIS background is observed,  which, moreover,  is compatible 
with the instanton-signal, it can not be claimed to be significant
given the uncertainty of the Monte Carlo simulations of the standard DIS background.
Therefore, only upper limits on the cross-section for QCD instanton-induced processes
are set, dependent on the kinematic domain considered~\cite{Adloff:2002ph}. From this 
analysis one may infer, via the saddle point correspondence, 
that the cross-section calculated within instanton-perturbation theory 
is ruled out for $(R/\rho)_\ast\lwig 0.84$, in a range $0.31\ {\rm fm}\,\lwig \rho_\ast\lwig\,0.33$ fm 
of effective instanton sizes. 
One should note, however, that 
in the corresponding -- with present statistics accessible -- kinematical range 
the running coupling is quite large, $\alpha_s(\rho^{-1}_\ast)\approx 0.4$, 
and one is therefore not 
very sensitive\footnote{This is of course welcome for the QCD-instanton searches at HERA, because 
it makes predictions  for the bulk of data quite reliable.} to the 
$I\bar I$-interaction, 
which appears in the exponent with coefficient $4\pi/\alpha_s\approx 31$. This should be contrasted with QFD, which
is extremely sensitive to $\Omega$, since $4\pi/\alpha_W\approx 372$.       
An extension of the present H1 limit on $(R/\rho)_\ast$ towards smaller $\rho_\ast$ and  
$\alpha_s(\rho_\ast^{-1})$, which should
be possible with increased statistics at HERA II, would be very welcome.
At present, the data do not exclude the cross-section predicted by instanton-perturbation theory 
for small $(R/\rho )_\ast \gwig 0.5$, as long as one probes only very small 
instanton-sizes $\rho_\ast\ll 0.3$~fm.

{\em 4.}
Finally, let us present the result of a state of the art evaluation of the cross-section~(\ref{gencross})
for QFD, including all the prefactors -- an analogous evaluation has been presented for DIS
in QCD in Ref.~\cite{Ringwald:1998ek}. For the case of fermion-fermion scattering via 
QFD instantons/sphalerons, as relevant at VLHC at the parton level, we find  
\begin{eqnarray}
\label{cross-qfd}
M_W^2\,\hat\sigma_{\rm ff}^{(I_W)} 
&=& \frac{\pi^{15/2}}{128}\,d_{\overline{\rm MS}}^2
\,
\left( \frac{4\pi}{\alpha_W(\mu )}\right)^{7/2}
\,
\left( \mu \rho_\ast \right)^{2\beta_0
} 
\\[1.5ex] \nonumber && \times
\left(\omega (\xi_\ast)\right)^{10}
\frac{1}{\sqrt{ 
\xi_\ast 
\left(\frac{\partial^2}{\partial\xi_\ast^2}\Omega_g(1,\xi_\ast )\right)
\left(\frac{\partial}{\partial\xi_\ast}\Omega_g(1,\xi_\ast )\right)^3
\left( \Omega_1(\xi_\ast)+2\,\Omega_2(\xi_\ast )\right)^{3}
} }
\\[1.5ex] \nonumber && \times
\exp\left[ -\frac{4\pi}{\alpha_W(\mu)}\,\left( 1+\Omega_g (1,\xi_\ast )
-(\xi_\ast -2 )\,
\frac{\partial}{\partial\xi_\ast} \Omega_g (1,\xi_\ast )
\right)\right]_{\left| \xi_\ast =2+\left(\frac{R}{\rho}\right)_\ast^2\right.}
\,,
\end{eqnarray}
expressed entirely in terms of the solutions of the saddle-point equations~(\ref{saddle-coll}).
The various factors in Eq.~(\ref{cross-qfd}) can be easily understood. The ones in the first 
line are mainly due to the 
square of the size distribution~(\ref{size-dist}), taken at the saddle-point. 
The power of $(4\pi/\alpha_W)$ is reduced
here from nominally $4\,N_c=8$ to $7/2$, because there are effectively $9$ saddle-point integrals 
giving rise -- 
apart from the square-root factor in the second line of Eq.~(\ref{cross-qfd}) -- 
 to a factor of $(4\pi/\alpha_W)^{-9/2}$. The explicit factor of $1/(2\,p_1\cdot p_2)=1/\hat s$ 
in Eq.~(\ref{gencross}) does not appear in Eq.~(\ref{cross-qfd}), because it is cancelled
by another explicit energy dependence in the factor 
${{\mathcal P}_{\rm ff}^W}_{\mid \ast}=8\pi^4\rho_\ast^6\,\hat s$. 
Finally, the last line in Eq.~(\ref{cross-qfd}) is just the main exponential~(\ref{gen-cross-holy}), 
${\rm e}^{-\Gamma_\ast}$. 

The prediction\footnote{\label{guess}At $\epsilon\sim 1$ it should be rather called an 
educated extrapolation or guess.}~(\ref{cross-qfd}) 
for the QFD instanton-induced fermion-fermion cross-section is 
displayed in Fig.~\ref{saddle-holy} (bottom) as a function of the scaled fermion-fermion cm energy
$\epsilon =\sqrt{\hat s}/(4\pi M_W/\alpha_W)$, for a choice $\mu = M_W$ 
of the renormalization scale. 
In the strict region of instanton-perturbation theory, 
$\epsilon\ll 1$, 
the cross-section is really tiny, e.g. $\hat\sigma_{\rm ff}^{(I_W)}\approx 10^{-141}$~pb at 
$\epsilon\approx 0.1$, but steeply growing. 
Nevertheless, it is expected to be 
unobservably small, $\hat\sigma_{\rm ff}^{(I_W)}\lwig 10^{-26}$~pb for $\epsilon\,\lwig\, 0.75$, 
in the conservative fiducial
kinematical region corresponding to $(R/\rho)_\ast\gwig 1$ 
inferred via the QFD--QCD analogy 
from lattice data and HERA. If we allow, however, for a slight extrapolation towards
smaller $(R/\rho)_\ast\approx 0.7$ -- still compatible with lattice results and HERA -- 
the prediction$^{\ref{guess}}$ rises to $\hat\sigma_{\rm ff}^{(I_W)}\approx 10^{-6}$~pb at $\epsilon\approx 1$, 
corresponding to a parton-parton cm energy 
of about $30$~TeV. In this case, QFD instanton-induced $B+L$ violating events will have 
observable rates at VLHC, which has a projected proton-proton cm energy of $\sqrt{s}=200$~TeV  
and a luminosity of about ${\mathcal L}\approx 6\cdot 10^5$~pb$^{-1}$\,yr$^{-1}$~\cite{vlhc},  
and an exciting phenomenology will emerge~\cite{Farrar:1990vb
}.
If we assume the prediction$^{\ref{guess}}$~(\ref{cross-qfd}) to be valid even at higher energies, corresponding
to even smaller $(R/\rho)_\ast$, than we can expect to be able to see the first signs 
of electroweak sphaleron production in present day or near future cosmic ray facilities and 
neutrino telescopes~\cite{Morris:1991bb
}, even before the commissioning of VLHC. In the meantime, we can try to improve our 
knowledge about QCD instantons on the lattice and in deep-inelastic scattering at HERA, 
with important implications also for QFD instantons at very high energies.  

\section*{Acknowledgements}
I would like to thank F.~Schrempp for many fruitful discussions and a careful
reading of the manuscript. Furthermore, I would like to thank the organizers of the
26th Johns Hopkins Workshop on Current Problems in Particle Theory, Heidelberg, 
August 1-3, 2002, in particular O.~Nachtmann, for inspiration and encouragement 
of the present work.

\end{document}